\documentclass[aps,prd,preprint,superscriptaddress,showpacs]{revtex4}
\usepackage{slashed}
\usepackage{graphicx}
\usepackage{verbatim}
\usepackage{ulem}
\usepackage{color}
\definecolor{My_red}        {cmyk}{0.00,1.00,1.00,0.20}

\newcommand{\bmat}{\left(\begin{array}}
\newcommand{\emat}{\end{array}\right)}
\newcommand{\beq}{\begin{equation}}
\newcommand{\eeq}{\end{equation}}
\newcommand{\wt}{\widetilde}

\def\ra{\rightarrow}

\def\Ld{\Lambda}
\def\ld{\lambda}
\def\f{\frac}

\def\bwt{\begin{widetext}}
\def\ewt{\end{widetext}}
\def\be{\begin{equation}}
\def\ee{\end{equation}}
\def\bea{\begin{eqnarray}}
\def\eea{\end{eqnarray}}
\def\bean{\begin{eqnarray*}}
\def\eean{\end{eqnarray*}}
\def\bary{\begin{array}}
\def\eary{\end{array}}
\def\bit{\begin{itemize}}
\def\eit{\end{itemize}}

\def\ra{\rightarrow}

\def\Ld{\Lambda}

\def\ld{\lambda}

\def\su5u1{SU(5) \times U(1)}
\def\fsu5u1{SU(5) \times U(1)'}
\def\so10{SO(10)}
\def\sq20{SO(10) \times SO(10)}

\def\ra{\rightarrow}

\def\Ld{\Lambda}
\def\ld{\lambda}
\def\f{\frac}

\def\L{\left(}
\def\R{\right)}

\def\ra{\rightarrow}

\def\Ld{\Lambda}

\def\ld{\lambda}

\def\su5u1{SU(5) \times U(1)}
\def\fsu5u1{SU(5) \times U(1)'}
\def\so10{SO(10)}
\def\sq20{SO(10) \times SO(10)}

\usepackage[centertags]{amsmath}
\usepackage{amssymb}

\begin{document}

\title{Highlights of Supersymmetric Hypercharge $\pm1$ Triplets }

\author{Zhaofeng Kang}
\email{zhaofengkang@gmail.com} \affiliation{Center for High-Energy
Physics, Peking University, Beijing, 100871, P. R. China}

\author{Yandong Liu}
\email{ydliu@itp.ac.cn} \affiliation{ State Key Laboratory of
Theoretical Physics,
      Institute of Theoretical Physics, Chinese Academy of Sciences,
Beijing 100190, P. R. China }

\author{Guo-Zhu Ning}
\email{ngz@mail.nankai.edu.cn} \affiliation{Center for High-Energy
Physics, Peking University, Beijing, 100871, P. R. China}

\date{\today}

\begin{abstract}

The discovery of a standard model (SM)-like Higgs boson with a
relatively heavy mass $m_h$ and hints of di-photon excess has deep
implication to supersymmetric standard models (SSMs). We consider
the SSM extended with hypercharge $\pm1$ triplets, and investigate
two scenarios of it: (A) Triplets significantly couple to the Higgs
doublets, which can substantially raise $m_h$ and simultaneously
enhance the Higgs to di-photon rate via light chargino loops; (B)
Oppositely, these couplings are quite weak and thus $m_h$ can not be
raised. But the doubly-charged Higgs bosons, owing to the gauge
group structure, naturally interprets why there is an excess rather
than a deficient of Higgs to di-photon rate. Additionally, the
pseudo Dirac triplet fermion is an inelastic non-thermal dark matter
candidate. Light doubly-charged particles, especially the
doubly-charged Higgs boson around 100 GeV in scenario B, are
predicted. We give a preliminary discussion on their search at the
LHC.

\end{abstract}


\maketitle

\section{Introduction and motivations}

The CMS and ATLAS collaborations discovered a new resonance around
126 GeV~\cite{Higgs:126}, the putative long-sought Higgs boson
predicted by the standard model (SM). This good news consolidates
supersymmetry (SUSY), which elegantly solves the gauge hierarchy
problem relating with Higgs (assumed to be a fundamental spin-0
particle), as the leading candidate for new physics. Nevertheless,
when we explain the Higgs boson in the popular minimal
supersymmetric SM (MSSM), two problems will arise. One is that
obtaining the relatively heavy mass ($m_h\simeq$126 GeV) renders a
quite serious fine-tuning~\cite{Kang:2012sy}. The other one is that
the hinted di-photon signal excess~\cite{Higgs:126} can not be
naturally understood. Therefore, naturally addressing these two
problems simultaneously, despite the waiting-for-confirmation for
the latter, gives us illuminating guide to go beyond the MSSM
(BMSSM).

Actually, the relatively heavy Higgs boson mass alone may open a
window for new model building. As been well known, in the MSSM the
tree-level mass of the SM-like Higgs boson mass lies below the
$Z-$boson mass, owing to the fact that the strength of Higgs quartic
coupling is determined by the electroweak gauge couplings. New
models thus should provide a significant Higgs quartic coupling
either at tree- or loop-level. A well known example is the singlet
extended MSSM such as the next-to MSSM (NMSSM). It possesses a large
coupling between the singlet and Higgs doublets, which can give an
additional Higgs quartic coupling at tree level. On top of that, it
has a singlet-doublet mixing effect to raise
$m_h$~\cite{Kang:2012sy,Mixng:NMSSM} (The related collider search
please see Ref.~\cite{Kang:2013rj}). Recently, other explorations on
BMSSM along the line of raising $m_h$ include: (A) The gauge group
extension, which gives non-decoupling $D-$terms~\cite{Hirsch:2012kv}
to lift $m_h$ (It requires the Higgs doublets to be effectively
charged under the new gauge group); (B) Vector-like particles
extension, which raises the Higgs boson mass at
loop-level~\cite{Ajaib:2012eb}; (C) $SU(2)_L$ triplet $T_0$ with
hypercharge 0 extension~\cite{Basak:2012bd,Delgado:2012sm}. $T_0$
couples to the Higgs doublets via $\ld_{T_0} {\rm Tr}H_u T_0H_d$
which, in raising $m_h$, is similar to the singlet-doublet coupling
in the NMSSM.
%except for the absence of mixing effect. 也可以有3-2混合！

However, when we further take into consideration the global fit on
Higgs signal strength data~\cite{Carmi:2012in}, most of the
aforementioned BMSSMs will be disfavored. In light of the best
global fit, we may need new particles carrying electric charges
which can directly enhance the decay width of Higgs to
di-photon~\cite{Carmi:2012in} and at the same time do not
appreciably affect other partial decay widths, especially
${\Gamma}(h\ra ZZ/WW)$. In the NMSSM, the light chargino, predicted
by natural SUSY, might be a candidate of such charged particles.
However, it only works when the singlet-doublet mixing is well
tempered~\cite{Choi:2012he}, which may introduce an extra
fine-tuning. The triplet $T_0$ extension may improve that case, by
virtue of new charged fermions from $T_0$. Indeed, they can readily
enhance the di-photon rate, given a large enough
$\ld_{T_0}$~\cite{Delgado:2012sm}.

In this paper we consider the low energy SUSY incorporating
$SU(2)_L$ triplets $T_{u,d}$, which carry hypercharges $\pm1$,
respectively. Similar to Ref.~\cite{Delgado:2012sm}, this BSSM is a
good case following the guide mentioned at the beginning. As been
noticed long ago by~\cite{Espinosa:1991gr} and more recently
by~\cite{Dine:2007xi,Agashe:2011ia,FileviezPerez:2012ab}, if
$T_{u,d}$ couple to the Higgs doublets with significant strengths,
the Higgs mass can be substantially enhanced. Simultaneously, the
singly-charged charginos, the mixture between these from $T_{u,d}$
and from the MSSM, are capable of enhancing the di-photon rate.
Different to Ref.~\cite{Delgado:2012sm}, the BMSSM considered in
this paper has doubly-charged Higgs bosons. When $T_{u,d}$ weakly
couple to the Higgs doublets, they can be naturally light and then
be viable candidates of new charged particles (But this scenario
fails in raising $m_h$). Besides, it may provide a pseudo Dirac
fermion to be an inelastic non-thermal dark matter (DM) candidate.

Either in the significant or weak coupling limit, light
doubly-charged particles are expected. Especially, the latter
definitely predicts light doubly-charged Higgs bosons around 100
GeV. They may have promising LHC discovery prospects.

%Finally, both scenarios predict

This paper is organized as follows. In Section II we study the MSSM
extended with hypercharge $\pm1$ triplets, two distinct scenarios
are explored and we stress their phenomenological highlights for
Higgs data and dark matter. Corresponding LHC search strategy and
prospects are also briefly commented. Discussion and conclusion are
casted in Section III and some necessary and complementary details
are given in the Appendices.

\section{Highlights of SUSY with Hypercharge $\pm1$ triplets}\label{LHiggs}

Motivated by naturally explaining the recent Higgs data,
including the relatively heavy SM-like Higgs
boson mass and/or the hinted excess of Higgs to di-photon rate, we
consider $SU(2)_L$ triplets $T_{u,d}$ extended SSMs (TSSMs). Here
$T_{u,d}$ carry hypercharges $-1$ and $+1$, respectively. Note that
both of them are needed for the sake of anomaly cancelation.

\subsection{Model setup}\label{setup}

The most relevant superpotential of the TSSM and the corresponding soft
SUSY-breaking Lagrangian are given by
\begin{align}
W_{H}&=\mu H_u\cdot H_d+\mu_T{\rm Tr}(T_u  T_d)+\ld_uH_u\cdot
T_uH_u+\ld_dH_d\cdot T_dH_d,\\
{\cal L}_{soft}&=\sum_{\Phi=H_{u,d},T_{u,d}}m_\Phi^2|\Phi|^2+ \L
B_\mu H_u\cdot H_d+B_{\mu_T}{\rm Tr}(T_u  T_d)+c.c.\R\cr &+\L
A_uH_u\cdot T_uH_u+A_dH_d\cdot T_dH_d+c.c.\R.
\end{align}
For simplicity, here all parameters are assumed to be real
(Incorporating CP violation will bring about quite different Higgs
phenomenologies~\cite{Balazs:2012bx}.). $T_u$ and $T_d$, following
the notation of Ref.~\cite{Espinosa:1991gr,Agashe:2011ia}, are
respectively written as
\begin{align}
T_u\equiv T_u^a\sigma^a=\left(\begin{array}{ccc}
\f{T_u^-}{\sqrt{2}}&-T_u^0\\
T_u^{--}&-\f{T_u^-}{\sqrt{2}}
\end{array}\right),\quad T_d\equiv T_d^a\sigma^a=\left(\begin{array}{ccc}
\f{T_d^+}{\sqrt{2}}&-T_d^{++}\\
T_d^0&-\f{T_d^+}{\sqrt{2}}
            \end{array}\right),
\end{align}
with $\sigma^a$ the three Pauli matrices. As usual, the doublets are
written as $H_u=(H_u^+,H_u^0)$ and $H_d=(H_d^0,H_d^-)$. For
illustration, we write the superpotential terms in component form
\begin{align}\label{comp}
H_u\cdot
T_uH_u\equiv(H_u)_\alpha\epsilon^{\alpha\beta}(T_u)_\beta^{~\gamma}(H_u)_\gamma
=\sqrt{2}H_u^+H_u^0T_u^--(H_u^0)^2T_u^0-(H_u^+)^2T_u^{--},
\end{align}%\cdot 是而分量旋量点成定义。
where the antisymmetric tensor $\epsilon^{\alpha\beta}$ has entry
$\epsilon^{12}=-1$. %The above definition of spinor product is
%applied to other quantities.

If we want to produce the small neutrino masses, as in the type-II
seesaw mechanism~\cite{TIIss}, the following lepton number violating
operators are introduced:
\begin{align}
W_{L}&=(\ld_{L})_{ij}L_i\cdot T_dL_j.
\end{align}
with $i,j=1,2,3$ the family indices. Then, the tiny neutrino masses
are the products of the small Yukawa couplings and triplet's vacuum
expected value (VEV):
$(m_{\nu})_{ij}=(\ld_L)_{ij}v_{T_d}\sim10^{-10}$ GeV. One would find
that the triplets' VEV of interest should be around the GeV scale,
so the Yukawa couplings are extremely small. Then, alternatively,
one may forbid such couplings and turn to the canonical seesaw
mechanism. In any case, this aspect of the TSSM is not of our main
concern. But we would like to stress that, the TSSM under
consideration can be embedded into the supersymmetric type-II seesaw
mechanism~\cite{Hambye:2000ui} without violating $R-$parity. In
other words, the lightest sparticle {LSP} DM hypothesis of SUSY can
be maintained here. In contrast, the models with hypercharge 0
triplets can not be embedded into the type-III seesaw mechanism and
meantime respect the $R-$parity.

\begin{figure}
\begin{center}
\includegraphics[scale=1]{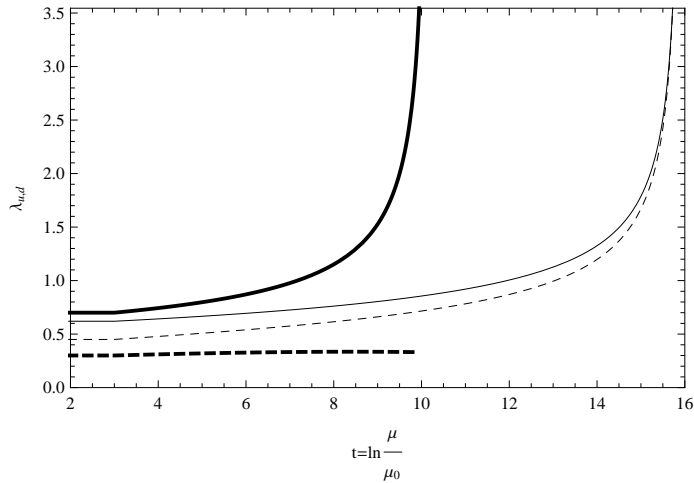}
\caption{\label{running} The plot of $\lambda_{u,d}$ verses the
running scale. Two groups of boundary values at the low energy are
chosen: (A) $\ld_u=0.3$ (thick dashed line) and $\ld_d=0.7$ (thick
line); (B) $\ld_u=0.45$ (thin dashed line) and $\ld_d=0.62$ (thin
line), showing the maximal $\ld_{u,d}$ endured by perturbativity up
to the GUT-scale.}
\end{center}
\end{figure}

In the TSSM, $\ld_u$ and $\ld_d$ are closely related to the ensuing
discussions. In one scenario studied later they will be required to
be large, so we want to know the scale at which perturbitivity
breaks. The relevant renormalization group equations (RGEs) are
casted in Appendix.~\ref{RGEs}. From it one can find, due to $h_t\gg
h_b$, the beta functions of $\ld_u$ and $\ld_d$ are asymmetric.
$\ld_d$ is favored to be moderately larger than $\ld_u$.
Fig.~\ref{running} shows that if they do not hit the Landau pole
below the GUT-scale, the maximal $\ld_d$ and $\ld_u$ are about 0.62
and 0.45, respectively. For $\ld_d=0.7$, it hits the Landau pole at
a scale $\sim10^{10}$ GeV. The implication to GUT is beyond the
scope of this work and we comment on that in the discussion.

In the following subsections we will turn our attention to the
highlights of the TSSMs, and focus on phenomenologies involving the
Higgs and DM. Two different scenarios according to the magnitudes of
$\ld_{u,d}$ are separately investigated.

\subsection{Scenario A: Large $\ld_{u,d}\sim 1$}

This scenario takes the advantage in rasing the SM-like Higgs boson
mass and the Higgs to di-photon rate simultaneously. However, they
tend to show a tension and fortunately the mixing effect can
relax it.

\subsubsection{Lifting the Higgs boson mass and pull-down mixing effect}

The CP-even Higgs boson sector consists of four states, with two
from the Higgs doublets and the rest from the Higgs triplets
$T_{u,d}^0$. Mixings between Higgs doubets and triplets are induced
by the triplets' VEV, which should be no more than a few GeVs due to
the bound from the $\rho$ parameter~\cite{Agashe:2011ia}.
Concretely, the triplets' VEV can be approximately determined in an
analytical way:
\begin{align}\label{vTud}
v_{T_{u,d}}\equiv&\langle T_{u,d}^0\rangle \simeq\f{ v^2}{m_{T_{u,d}}^2+\mu_T^2}{\cal M}_{u,d},\cr
{\cal M}_{u}\equiv& A_u\sin^2\beta -\ld_d \mu_T\cos^2\beta-\ld_u\mu\sin2\beta,\cr
{\cal M}_{d}\equiv& A_d\cos^2\beta -\ld_u \mu_T\sin^2\beta-\ld_d\mu\sin2\beta,
\end{align}
with $\tan\beta=\langle H_u^0\rangle/\langle H_d^0\rangle$. These
approximations are valid only for the case with heavy scalar
triplets and a not anomalously large $B_{\mu_T}$. It is clearly seen
that $\langle T_{u,d}^0\rangle\lesssim 1$ GeV, given a small
$\mu_T\sim 100$ GeV and large $\ld_{u,d}\sim 1$, requires quite
heavy triplets: $m_{T_{u,d}}\gtrsim{\cal O}(1)$ TeV. Otherwise, we
have to greatly tune parameters to make ${\cal M}_{u,d}$ lie around
the GeV scale.

On the other hand, $m_{T_{u}}$ can not be too heavy when naturalness
are taken into account. $H_u$ couples to $T_u$ with a large Yukawa
coupling, so $m_{H_u}^2$ receives a large negative correction
proportional to $m_{T_u}^2$, which, in the leading logarithmic
approximation, can be estimated by means of the renomalization group
equation (RGE):
\begin{align}
16\pi^2\f {d m_{H_u}^2}{dt}=6h_t^2\L m_{Q_3}^2+m_{\wt U^c}^2\R+12\ld_u^2 m_{T_u}^2+...
\end{align}
This allows us to derive an upper bound ($m_{T_d}^2$ can be well
beyond this bound since $T_d$ does not couple to $H_u$, but we do
not consider such a hierarchy here.):
\begin{align}
m_{T_u}\lesssim1.2\times\L\f{F^{1/2}}{10}\R\L\f{0.5}{\ld_u}\R\L\f{30}{\log \Ld/{\rm TeV}}\R^{1/2}\,\rm TeV,
\end{align}
with $F$ the degree of fine-tuning and $\Ld$ the SUSY-breaking
mediation scale. In the above estimation $\Ld$ is set equal to the
grand unification theory (GUT)-scale. But even if we take $\Ld=10^6$
GeV, the upper bound is merely doubled. In other words, in the large
$\ld_{u,d}$ scenario, naturalness does not permit the triplet VEVs
(at least $v_{T_u}$) to be far below one GeV. This has important
implications to the mixing effect.

We first investigate the TSSM specific effect on raising the SM-like
Higgs boson mass $m_h$ without considering the mixing effect. It is
convenient to employ the field decomposition as
\begin{align}\label{GSB:basis}
H_u^0=&v_u+\f{1}{\sqrt{2}}(h_1\cos\beta+h_2\sin\beta)+\f{i}{\sqrt{2}}(P_1\cos\beta+G^0\sin\beta),\cr
H_d^0=&v_d+\f{1}{\sqrt{2}}(-h_1\sin\beta+h_2\cos\beta)+\f{i}{\sqrt{2}}(P_1\sin\beta-G^0\cos\beta),\cr
T_u^0=&v_{T_u}+\f{1}{\sqrt{2}}\L h_3+iP_2\R,\quad T_d^0=v_{T_d}+\f{1}{\sqrt{2}}\L h_4+iP_3\R,
\end{align}%这个基下可以看出$S_2$差不多就是SM-like的Higgs。
where $G^0$ is the Goldstone boson and $P_i$ are the CP-odd Higgs
bosons. The CP-even Higgs boson mass square matrix ${M}_{S}^2$, in
the basis $(h_1,h_2,h_3,h_4)$, is given by Eq.~(\ref{ms2}).
Neglecting both the doublet-doublet and doublet-triplet mixing
effects for the time being, we then get $m_h^2$, which is
approximated by
\begin{align}\label{upper}
({M}_{S}^2)_{22}=&m_Z^2\left[\cos^22\beta+\f{4}{g^2}\L\ld_d^2\cos^4\beta+\ld_u^2\sin^4\beta\R\right],
\end{align}
with $m_Z^2=g^2v^2$ and $g^2=(g_1^2+g_2^2)/2$. The new contributions
originate from $\ld_{u,d}H_{u,d}T_{u,d}H_{u,d}$, which give new
quartic terms, e.g., $|F_{T_u^0}|\supset \ld_u^2|H_u^0|^4$.
Noticeably, provided that $\ld_{u,d}$ are order one coupling
constants, the tree-level Higgs boson mass will get a considerable
enhancement, regardless of the value of $\tan\beta$.

Such an interesting feature, obviously, is attributed to the fact
that the quartic terms $|H_d^4|$ and $|H_u^4|$ are generated
simultaneously. To our knowledge, basically the $m_h$ dependence on
the angle $\beta$ can be classified into three types: (I) The
MSSM-like case where the quartic term is determined by the
vector-like $D-$terms (Namely $H_{u,d}$ form a vector-like
representation under the gauge group, such as $SU(2)_L\times
U(1)_Y$), and takes the form of $(|H_u^0|^2-|H_d^0|^2)^2\propto
\cos^22\beta$. A large $\tan\beta$ is then required to give a heavy
$m_h$; (II) The NMSSM-like case where the quartic term is
$|H_u^0H_d^0|^2\propto \sin^22\beta$ and thus a small
$\tan\beta\sim1$ is required to lift $m_h$. The SSMs with
hypercharge-0 triplets~\cite{Basak:2012bd,Delgado:2012sm} also fall
into this category; (III) The TSSMs considered in this work as well
as the models extended by a new gauge group under which $H_u$ and
$H_d$ form a chiral representation. As will be explicitly seen, the
Higgs to di-photon rate excess also has a strong dependence on
$\beta$. Therefore, the above classification provides a guidance to
build models which can simultaneously lift $m_h$ and the Higgs to
di-photon rate.

We now take the mixing effect into account, which may reduce the
maximal enhancement given in Eq.~(\ref{upper}). To see that, we
consider the sub-matrix of $M_S^2$, consisting of the entries
involving states 2 and 3 (namely only including the mixing with
$T_u^0$). From Eq.~(\ref{ms2}), it is not difficult to find that
this $2\times2$ matrix shows the relation
\begin{align}\label{}
({M}_{S}^2)_{22}({M}_{S}^2)_{33}\sim {\cal
O}(0.1)({M}_{S}^2)_{23}^2.
\end{align}
Thereby, the mixing effect can be appreciable and thus pulls-down
the lighter eigenvalue indicated by Eq.~(\ref{upper}), with an
amount estimated to be~\cite{Kang:2012sy}
\begin{align}\label{}
\Delta m_h\approx&
-\f{1}{2}\sqrt{({M}^2_{S})_{22}}\f{({M}_{S}^2)_{23}^4}{({M}_{S}^2)_{33}({M}_{S}^2)_{22}}\cr
=&-2\L\f{{\cal M}_u^2}{m_{T_u}^2}\R\L
\f{m_h}{\sqrt{({M}^2_{S})_{22}}}\R\L \f{v^2}{m_h}\R\rm\,GeV.
\end{align}
The mixing with $T_d^0$ similarly contributes to the pulling-down
effect and then roughly the above estimation should be doubled.

\begin{figure}[htb]
\begin{center}
\includegraphics[width=3in]{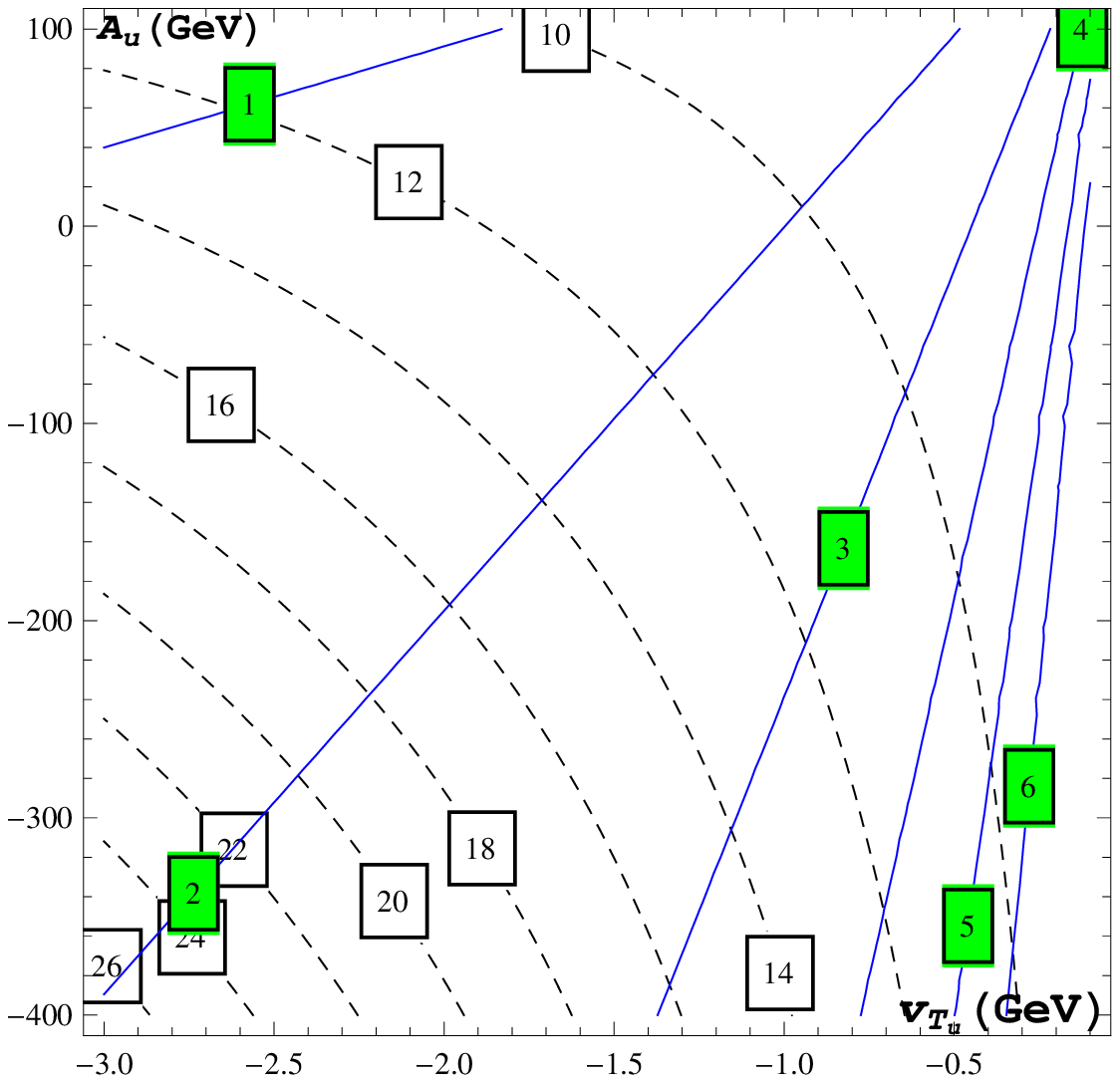}~~~~
\includegraphics[width=2.95in]{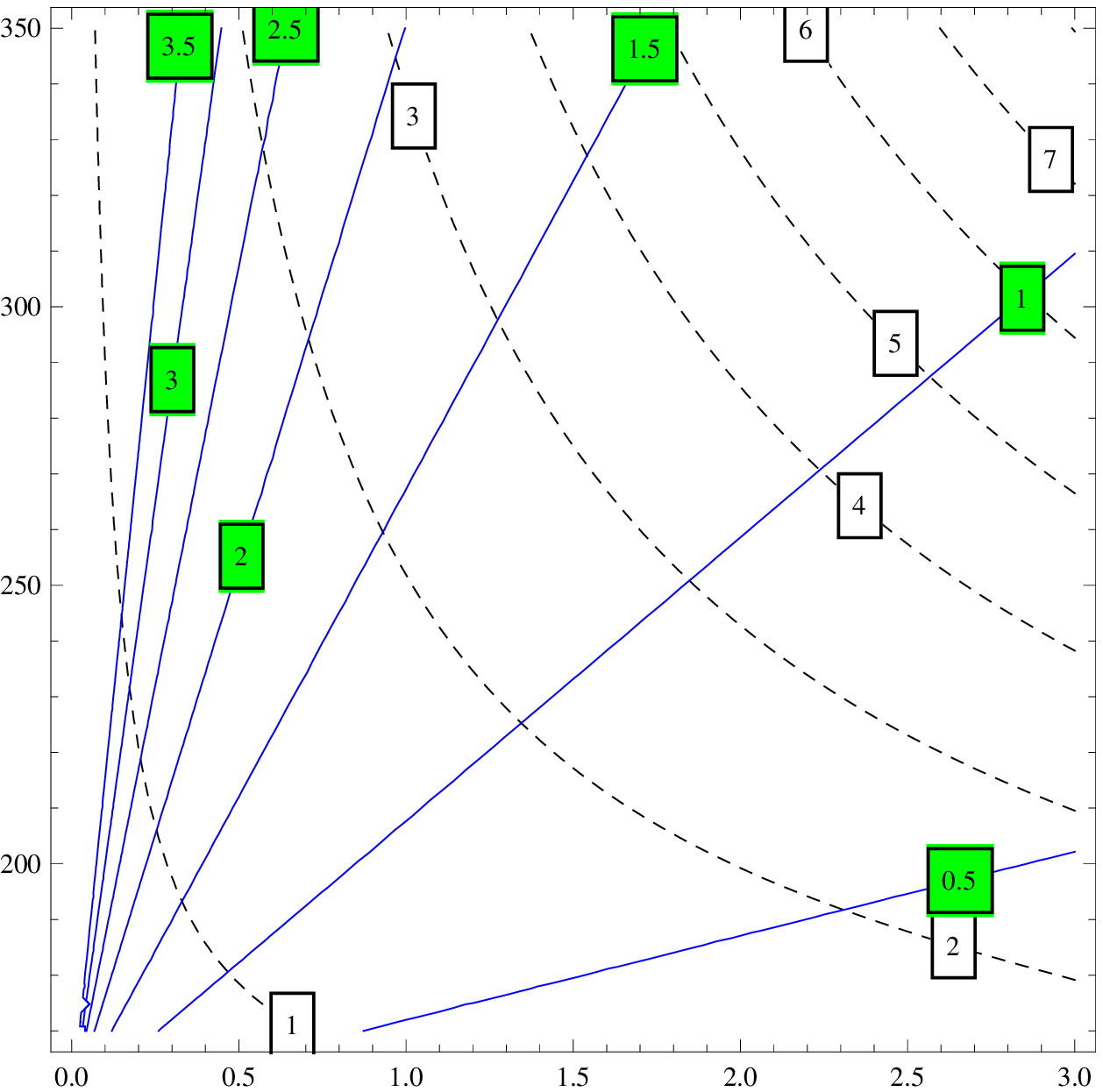}
\end{center}
\caption{\label{mh} Contour plots of the amount of Higgs boson mass
reduction (in GeV, dashed lines with white labels) from the
doublet-triplet pulling-down mixing effect and the triplet soft mass
$m_{T_{u}}$ (in TeV, solid lines with green labels) on the
$v_{T_u}-A_u$ plane. Left panel: Large reduction by setting
$\ld_d=0.70$, $\ld_u=0.45$, $v_{T_d}=-2.0$ GeV and $A_d=-300$ GeV.
Right panel: Small reduction by setting $\ld_d=0.70$, $\ld_u=0.30$,
$v_{T_d}=-0.5$ GeV and $A_d=200$ GeV. Other not much relevant
parameters are common to both cases: $\mu=200$ GeV, $\mu_T=250$ GeV,
$B\mu=400\times200$ GeV$^2$, $B\mu_T=200^2$ GeV$^2$ and
$\tan\beta=1.5$ (Actually, effect of $\mu$, $\mu_T$ and $\tan\beta$
can be absorbed into $A_{u,d}$, see Eq.~(\ref{vTud})).}
\end{figure}

The pulling-down mixing effect is tunable. Through a mild tuning one
can obtain relatively small ${\cal M}_{u,d}$, which helps to not
only reduce $v_{T_{u,d}}$ but also weaken the pulling-down effect.
This is consistent with the $\rho$ parameter constraint. However,
sometimes the new quartic terms excessively enhance $m_h$, which
will be the case when we want to further account for the di-photon
excess, then we need relatively large ${\cal M}_{u,d}$, or
equivalently $v_{T_{u,d}}$, to strength the pulling-down effect.
Then, from Eq.~(\ref{vTud}) it is known that negative $A_{u,d}$ are
favored. For illustration, we plot these two limiting cases of
mixing effect on Fig.~\ref{mh}.
% if we

\subsubsection{Rasing the Higgs to di-photon rate via charginos}

The Higgs to di-photon rate $R_{\gamma\gamma}$ can probe charged
particles beyond the SM. As usual, the rate is defined as
\begin{align}\label{}
R_{\gamma\gamma}\equiv\f{\sigma(gg\ra h){{\rm Br}(h\ra \gamma\gamma)
}}{\sigma_{\rm SM}(gg\ra h){{\rm Br_{\rm SM}}(h\ra \gamma\gamma)}}.
\end{align}
For a generic particle content, the decay width of Higgs to
di-photon is formulated as~\cite{Carena:2012xa}
\begin{align}\label{hgg:general}
\Gamma(h\ra\gamma\gamma)=\f{\alpha^2m_h^3}{1024\pi^3}\left|\f{g_{hVV}}{m_V^2}Q_V^2{\cal
A}_1(\tau_V)+ \f{2g_{hf\bar f}}{m_f}N_{c,f}Q_f^2{\cal
A}_{1/2}(\tau_f)+\f{g_{h|S|^2}}{m_S^2}N_{c,s}Q_S^2{\cal
A}_0(\tau_S)\right|^2,
\end{align}
with $N_c$ the color factor and $\tau_i\equiv 4m_i^2/m_h^2$. Here
$V,~f,$ and $S$ denote for a vector boson, Dirac fermion and charged
scalar, respectively. Their electric charges as well as couplings to
the Higgs boson are labeled as $Q_{V}$ and $g_{hVV}$, etc. In the
limits $\tau_i\gg 1$, the loop-functions ${\cal A}_i$ take
asymptotic values ${\cal A}_1\ra -7$, ${\cal A}_{1/2}\ra 4/3$ and
${\cal A}_0\ra 1/3$. Specified to the SM, the $W-$boson and top
quarks dominantly contribute to Eq.~(\ref{hgg:general}) and their
loop functions are
\begin{align}\label{SM:2g}
{\cal A}_1(\tau_W)\approx-8.34,\quad 3\times (2/3)^2{\cal
A}_{1/2}(\tau_t)\approx1.84.
\end{align}
To get it we have fixed $m_h=126$ GeV. In the TSSM, the
singly-charged charginos and doubly-charged Higgs bosons are two
promising candidates to make $R_{\gamma\gamma}>1$. The
doubly-charged charginos are irrelevant, which is due to the absence
of their direct couplings with the SM-like Higgs boson (See the
superpotential Eq.~(\ref{comp})). As for the doubly-charged Higgs
bosons, despite of their large electric charges and moreover large
couplings to the Higgs boson from D-term, are neither irrelevant.
The reason is that in the large $\ld_{u,d}$ scenario only quite
heavy $m_{T_{u,d}}^2$ are considered. But they will play important
roles in scenario B.

So, we only need to consider the contributions from the light
singly-charged charginos. For a mixed Driac system with a mass
matrix $M_F$, we have~\cite{Higg2g,Carena:2012xa}
\begin{align}\label{fermion}
\sum_i \f{2g_{hf_i\bar f_i}}{m_{f_i}}=
\f{1}{\sqrt{2}}\f{\partial}{\partial v}\log \L {\rm det} M_F^\dagger
M_F\R,
\end{align}
with $f_i$ the mass eigenstates having properly heavy eigenvalues.
Applying this formula to top quarks, we get $2g_{ht\bar
t}/m_t(=g_{hWW}/m_W^2)=\sqrt{2}/v$ with $v=174$ GeV. We now apply it
to the sinly-charged chargino system, which consists of three Dirac
fermions, winos $\wt W^\pm$, Higgsinos $\wt H_{u,d}^\pm$ as well as
triplinos $\wt T_{u,d}^\pm$. Their mass terms are given by
\begin{align}\label{}
(-i\wt W^-,\wt H_d^-,\wt T_u^-)\left(\begin{array}{ccc}
               M_2 & g_2 v_u & 0\\
                 g_2 v_d & \mu & \sqrt{2}\ld_d v_d\\
                 0 & \sqrt{2}\ld_u v_u & \mu_T
            \end{array}\right)\left(\begin{array}{c}
               -i\wt W^+\\
                 \wt H_u^+\\
                 \wt T_d^+
            \end{array}\right),
\end{align}
where terms proportional to $v_{T_{u,d}}$ have been safely
neglected. Denoting the charginos in the mass eigenstate as
$\chi_i\,(i=1,2,3)$, it is straightforward to calculate the
effective coupling defined in Eq.~(\ref{hgg:general}):
\begin{align}\label{hchichi}
\f{2g_{h\bar\chi\chi}}{m_{\chi}}\equiv\sum_{i=1}^3\f{2g_{h\bar\chi_i\chi_i}}{m_{\chi_i}}=\f{
{2\sin 2 \beta} \left(2  \lambda _d \lambda _u {v}/{\mu_T}+g_2^2
v/M_2\right)} {{\sin 2 \beta} \left(2  \lambda _d \lambda _u
{v}/{\mu_T}+g_2^2 v/M_2\right)-2 \mu/v}\f{\sqrt{2}}{v}.
\end{align}
Immediately, it is seen that a small $\tan\beta\sim1$, as in the
MSSM, is a necessary condition to enhance the Higgs to di-photon
rate by charginos~\footnote{This is a generic conclusion if the
Dirac fermions running in the loop develop a $v-$dependence only via
the mixing induced by electro-weak symmetry breaking.}. Actually,
from the counterpart of Eq.~(\ref{hchichi}) in the MSSM, we can get
Eq.~(\ref{hchichi}) through the shift $g_2^2 v/M_2\ra g_2^2 v/M_2+2
\lambda _d \lambda _u {v}/{\mu_T}$. Thus, in the decoupling limits
$\ld_{u,_d}\ll1$ or/and $\mu_T\gg v$, the triplinos are irrelevant
to the Higgs di-photon decay and the MSSM result is recovered. Then
the di-photon excess is no more than 30$\%$, even if we work in the
unrealistic limit with $\tan\beta=1$~\cite{Blum:2012ii}. Thus, the
case with non-decoupling $\wt T_{u,d}$ is of interest.

With Eq.~(\ref{hchichi}), the signature strength of Higgs to
di-photon is given by
\begin{align}\label{RGG}
R_{\gamma\gamma}\approx\left|1-\f{1.84}{8.34-1.84}\times
\f{3}{4}\f{v}{\sqrt{2}}\f{2g_{h\bar\chi\chi}}{m_{\chi}}\right|^2,
\end{align}
where we have used Eq.~(\ref{SM:2g}) and normalized the new physics
contribution to the top quark contribution, as indicated by the
second term in the above equation. Moreover, we assume that all
charginos' loop functions ${\cal A}_{1/2}(\tau_{\chi_i})$ are equal
to ${\cal A}_{1/2}(\tau_{t})$. But actually the lightest chargino
mass is about half of the top quark mass, and thus ${\cal
A}_{1/2}(\tau_{\chi_1})$ is slightly larger than ${\cal
A}_{1/2}(\tau_{t})$. As a result, Eq.~(\ref{RGG}) typically, but not
appreciably, underestimates the chargino contribution.

\begin{figure}[htb]
\begin{center}
\includegraphics[width=3in]{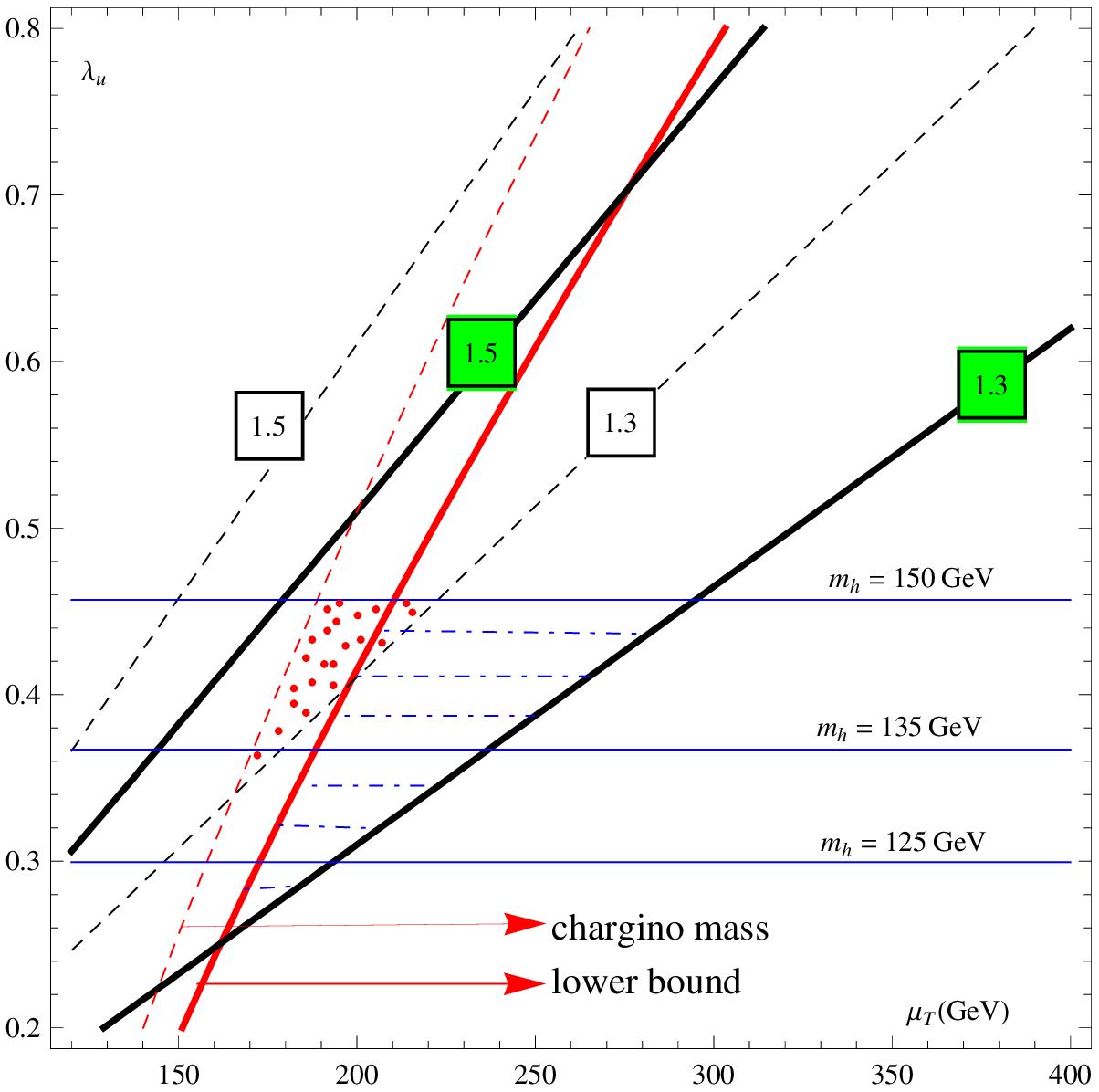}~~~~
\includegraphics[width=3in]{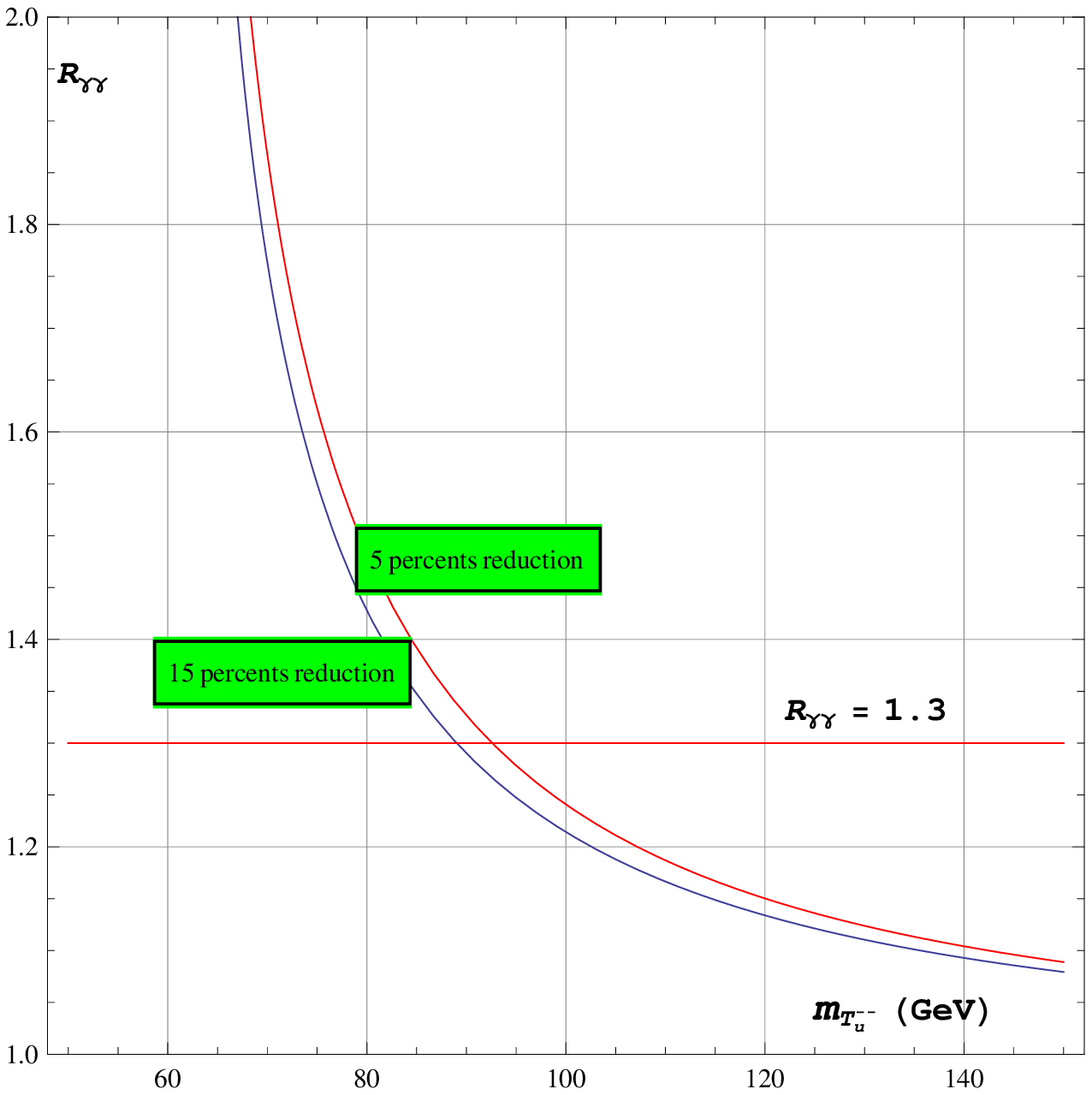}
\end{center}
\caption{\label{diphoton} %Plots of Higgs to di-photon signature
%strength.
Left panel (scenario A and on the $\mu_T-\ld_u$ plane): A
comparative study on the Higgs to di-photon signature strength
$R_{\gamma\gamma}$ with different wino mass $M_2$, setting
$\ld_d=0.7$, $\tan\beta=1.5$ and $\mu=200$ GeV. The tree-level
SM-like Higgs boson mass without considering pulling-down effect is
labeled by the horizontal lines. It is allowed to be mildly heavier
than 125 GeV. Two values of $M_2$ are taken: (I) $M_2=300$ GeV. The
charigno mass lower bound $m_{\chi_1}>94$ GeV~\cite{chargino:bound}
is labeled by the thick red line, and $R_{\gamma\gamma}=1.3
\,\,(1.5)$ are labeled by the thick black lines. The triangle filled
by blue dashed lines denote the allowed region with
$R_{\gamma\gamma}>1.3$; (II) $M_2=600$ GeV. $R_{\gamma\gamma}$ and
chargino mass are labeled by dashed lines. The triangle filled by
red points gives $R_{\gamma\gamma}>1.3$ and is significantly smaller
than the one in case I. Right panel (scenario B): A plot of
$R_{\gamma\gamma}$ varying with the doubly-charged Higgs boson mass,
and two cases with $5\%$ and $15\%$ reduction from singly-charged
Higgs boson are plotted, respectively. }
\end{figure}

The numerical results of $R_{\gamma\gamma}$ and the Higgs boson mass
are shown on the left panel of Fig.~\ref{diphoton}. We find that:
(I) The large di-photon rate may have a tension with $m_h=126$ GeV,
but the pulling-down effect can relax it; (II) Only the product
$\ld_{u}\ld_d$ is relevant to determine $R_{\gamma\gamma}$, so, in
light of the discussion in Section~\ref{setup}, we can take
$\ld_{u}=0.4$ and $\ld_d=0.62$ to get viable enhancements without
spoiling perturbativity below the GUT-scale; (III) All charginos,
including winos, should be properly light, which means that the MSSM
charginos have a non-negligible contribution to $R_{\gamma\gamma}$.
This contribution helps us to obtain a relatively large
$R_{\gamma\gamma}$ with $\ld_u\ld_d$ tolerated by perturbativity. To
end up this subsection, we mention that the sign of
$g_{h\bar\chi\chi}$, as in most models, is a result of parameter
tuning (or artificial choice). In the following we will present a
scenario to naturally understand the origin of the sign.

%10%是指振幅水平上的，没有模方。

\subsection{Scenario B: The small $\ld_{u,d}\ll 1$ limit}

We now switch to the scenario B characterized by very small
couplings $\ld_{u,d}\ll1$. Despite of the failure of raising $m_h$,
this scenario is still of great interest, by virtue of its elegant
explanation to the Higgs to di-photon excess. On top of that, this
scenario can provide an inelastic dark matter (DM) candidate.

\subsubsection{Rasing the di-photon rate via doubly-charged Higgs bosons}

As seen from Eq.~(\ref{vTud}), now the masses of doubly-charged
Higgs bosons can be light without spoiling the $\rho$ parameter.
Then, the charginos' contribution is reduced to the MSSM case, but
instead $H^{\pm\pm}$ is able to enhance $R_{\gamma\gamma}$.
Couplings between $H^{\pm\pm}$ and the Higgs boson come from the
$D-$term. The interesting point is that, by virtue of supersymmetry,
the couplings are fixed by the gauge group structure of the model
rather than input by hand. Explicitly, they are extracted from
Eq.~(\ref{Dterms}):
\begin{align}\label{hdoubly}
V_D\supset&
\f{g_2^2-g_1^2}{2}(|H_u^0|^2-|H_d^0|^2)\L|T_u^{--}|^2-|T_d^{++}|^2\R
-\f{g_1^2}{2} (|H_u^0|^2-|H_d^0|^2)\L|T_u^{-}|^2-|T_d^{+}|^2\R\cr =&
-\sqrt{2}\cos2\beta\cos
2\theta_w\f{m_Z^2}{v}h_2\L|T_u^{--}|^2-|T_d^{++}|^2\R\cr
&+\sqrt{2}\cos2\beta(\sin^2\theta_w/\cos
2\theta_w)\f{m_Z^2}{v}h_2\L|T_u^{-}|^2-|T_d^{+}|^2\R,
\end{align}%1202.6621v1中提及了对H++的约束，可以到100 GeV。$\cos2\theta_w\approx0.54$
where $\theta_w$ is the Weinberg angle with
$\sin^2\theta_w\approx0.23$.

We now show that we have a way to naturally explain
$R_{\gamma\gamma}>1$, rather than the other way around. From the
second line of Eq.~(\ref{hdoubly}) it is seen that, there is a
relative sign between the two terms in the bracket. It is a
consequence of $T_{u}^{--}$ lying at the 21-position in the $T_u$
matrix while $T_{d}^{++}$ at the 12-position in the $T_d$ matrix
(or, essentially, $T_u$ and $T_d$ carrying opposite hypercharges.).
Hence, if $T_d^{++}$ is very heavy and $T_u^{--}$ is sufficiently
light, say due to $m_{T_u}^2+\mu_T^2\ll m_{T_d}^2+\mu_T^2$ (Such a
hierarchy can be naturally realized in the gauge mediated
SUSY-breaking models by coupling $T_d$ to
messengers~\cite{Kang:2012ra}.), then $T_d^{++}$ decouples and we
have the effective coupling:
\begin{align}\label{H++}
\f{g_{h|T_u^{--}|^2}}{m_{T_u^{--}}^2}=\cos2\beta\cos
2\theta_w\f{m_Z^2}{m_{T_u^{--}}^2}\f{\sqrt{2}}{v}.
\end{align}
The effective coupling constant given by Eq.~(\ref{H++}) is
negative, so it elegantly explains why the doubly-charged-loop
constructively rather than destructively interferes with the
$W-$loop.

The enhancement of $R_{\gamma\gamma}$ is significant only in the
large $\tan\beta$ limit, and moreover it sensitively depends on the
doubly-charged Higgs boson mass. Concretely, in the large
$\tan\beta$ limit, we have
\begin{align}\label{}
R_{\gamma\gamma}\approx\left|1+\f{2.16}{8.34-1.84}\times\f{m_Z^2}{m_{T_d^{++}}^2}
{{\cal A}_0(\tau_{T_u^{--}})}+({T_u^{-}-\rm reduction})\right|^2,
\end{align}
which, actually, depends solely on the mass $m_{T_u^{--}}$. The
singly-charged Higgs boson with comparable mass reduces the
doubly-charged Higgs boson contribution by less than 10$\%$, which
will be regarded as a constant hereafter. The right panel of
Fig.~\ref{diphoton} clearly shows that an excess
$R_{\gamma\gamma}\gtrsim 20\%$ requires a doubly-charged Higgs boson
with mass below 105 GeV, even if the $T_u^{-}-$reduction is only
$5\%$.

Comments are in orders. First, although there is a great number of
works using a doubly-charged Higgs field to enhance the di-photon
rate~\cite{Arhrib:2011vc}, here we find that only supersymmetry can
naturally determine the sign of the effective coupling between the
SM-like Higgs boson and the doubly-charged Higgs boson. Second, from
the previous deduction it is tempting to conjecture that, this
sign-determination mechanism can be generalized to any SSMs having a
gauge group $G$ and some QED-charged chiral fields, which, along
with the Higgs doublets, form vector-like representations of $G$.
Third, the TSSM in scenario B has to give up the merit of lifting
$m_h$~\footnote{In principle, we can keep $\ld_u\sim{\cal O}(1)$ to
enhance $m_h$, but it is at the price of a big fine-tuning
($\sim1\%$) to obtain a small $v_{T_d}$.}, and we can enhance $m_h$
using methods reviewed in the introduction. But the method like in
the NMSSM, which requires a small $\tan\beta\sim 1$, may be
contradict with the di-photon rate enhancement.

\subsubsection{Pseudo-Dirac fermions as a non-thermal inelastic DM candidate}

In scenario B, the TSSM may provide an inelastic DM
candidate~\cite{Hall:1997ah,IDM}~\footnote{Actually,
Ref.~\cite{Chun:2012yt} has also discussed such an inelastic DM,
focusing on its aspect of indirect detection. We thank E. J. Chun
for pointing their earlier work, and moreover the PAMELA constraints
on the this DM, to us.}. Barring large cancelation, $\mu_T$ should
be around $m_{{T_d}^{++}}\simeq100$ GeV. Then, the neutral triplinos
$\wt T_{u,d}^0$ can be the lightest sparticles (LSP). They are
degenerate in mass and form a Dirac fermion in limits $\ld_{u,d}\ra
0$. However, the degeneracy is lift by the mixing with the MSSM
neutralinos, which are Majorana fermions. The corresponding mass
splitting is proportional to the small VEVs $v_{T_{u,d}}$, so it is
naturally small. As a consequence, the Dirac triplino becomes pseudo
Dirac.

We now study the mass splitting. In the basis $\chi^T_0=(-i\wt
B,-i\wt W^3,\wt H_d^0,\wt H_u^0,\wt T_d^0, \wt T^0_u)$, the
$6\times6$ neutralino mass matrix is
\begin{align}\label{}
M_{\chi^0}\approx\left(
\begin{array}{cccccc}
 M_1 & 0 & -\frac{ g_1   v_d }{\sqrt{2}} & \frac{ g_1   v_u  }{\sqrt{2}} & \sqrt{2}g_1v_{T_d} & -\sqrt{2}g_1v_{T_u} \\
   & M_2 & \frac{ g_2   v_d }{\sqrt{2}} & -\frac{ g_2   v_u  }{\sqrt{2}} & -\sqrt{2}g_2v_{T_d} & \sqrt{2}g_2v_{T_u} \\
   &   & 0 & -\mu  & \frac{ v_d  \ld_d}{\sqrt{2}} & 0 \\
 &  &  & 0 & 0 & \frac{ v_u   \ld_u }{\sqrt{2}} \\
   &   &  &   & 0 &  \mu_T  \\
   &   &   &   &   & 0
\end{array}
\right).
\end{align}
The analytical analysis is difficult, but we can calculate the mass
splitting in an approximate way. At tree-level, the splitting is
attributed to the mixing between the triplinos and MSSM neutralinos.
We consider a simplified case where mixing with wino dominates and
then obtain a $3\times3$ neutralino mass matrix:
\begin{align}\label{}
M_{\chi^0}\approx\left(
\begin{array}{ccc}
     M_2  & -\sqrt{2}g_2v_{T_d} & 0 \\
 -\sqrt{2}g_2v_{T_d}& 0 &  \mu_T  \\
 0   &  \mu_T  & 0
\end{array}
\right).
\end{align}
We have set $v_{T_u}\ll v_{T_d}$, which is a reasonable assumption
since $m_{T_u}^2\gg m_{T_d}^2$. The mass splitting can be calculated
to be
\begin{align}\label{delta}
\delta=\f{M_2}{M_2^2-\mu_T^2}(\sqrt{2}g_2v_{T_d})^2.
\end{align}
It is seen that, for a sub-GeV triplet VEV and moreover moderately
heavy $M_2\sim500$ GeV, $\delta$ is at the sub-MeV scale. If the
bino is considerably light, say $M_1\simeq M_2/3$ suggested by the
gaugino mass unification (Actually, one can include the bino
contribution by adding a similar term to Eq.~(\ref{delta}) with the
replacement $M_2\ra M_1$ and $g_2\ra g_1$), the above estimation is
only about half of the actual value. When $\mu$ is rather light, it
tends to reduce the splitting. In a word, the full result
accommodates a split over a rather wide range.

As been well known, DM participating in full electroweak (EW) gauge
interactions has a too large annihilation cross section, say $\sigma
v\sim g_2^4/16\pi \mu_T^2$, which then renders the DM relic density
far smaller than $\Omega_{\rm DM}h^2\sim0.1$. Although the heavy DM
around one TeV has a proper annihilation rate, it is not the case
considered here. We thus go beyond the thermal DM scenario. Consider
a highly bino-like (So it almost does not couple to the $Z-$boson)
ordinary LSP (OLSP), it acquires a proper thermal relic density and
late decays into the pseduo-Dirac fermion plus a pair of light
fermions. The decay is mediated by a slepton (See a similar scenario
in a different context~\cite{Kang:2010mh}.), with decay width
estimated to be
\begin{align}\label{}
    \Gamma(\wt B\ra \wt T_{u,d}^0  f f') \sim&\f{1}{(4\pi)^3}
    g_1^4 V_{14}^2 \f{M_{1}^5}{m_{\wt f}^4} P({M_1^2}/{\mu_{T}^2})\sim 10^{-18}  {\rm
    GeV}.
\end{align}
To get the final estimation, we have set the mixing angle between
bino and triplino $V_{14}\sim g_1^2 v_{T_d}^2/(\mu_T m_{\wt B})\sim
10^{-4}$, slepton mass at one TeV and the phase space factor
$P\sim10^{-2}$, which in principle can be even smaller in magnitude
provided sufficient degeneracy between bino and DM. Hence the decay
can happen after the DM freezing-out, and the bino can transfer its
number density into the DM.

A weak scale inelastic DM with sub-MeV mass splitting has interesting
phenomenological applications. First, it
provides a way to avoid the stringent spin-independent bound on the
$Z-$mediated DM-nucleon scattering, which makes the triplino-like DM
to be allowed by the DM direct detection
experiments~\cite{Hall:1997ah,DeRomeri:2012qd} like
XENON100~\cite{Aprile:2012nq} (It, along with the WMAP, raises question
on naturalness of the conventional neutralino LSP DM~\cite{Kang:2012sy}).
Additionally, it may be a candidate to explain the 511keV
line~\cite{Finkbeiner:2007kk} or the DAMA/LIBRA experiment result~\cite{IDM}.
It can also be used for some other purposes~\cite{Gao:2011ka}.

But indirect detection, such as the anti-proton flux measured by
PAMELA~\cite{PAMELA}, places a stringent bound on the triplino DM.
Ref.~\cite{Chun:2012yt} finds that its mass below $\sim$800 GeV may
have been excluded already. That renders light triplino DM in
scenario B problematic. But our main points discussed above are
still meaningful, because they can be used to make the heavier
inelastic triplino DM allowed by both PAMELA and XENON100.

\subsection{The preliminary of LHC search}

We have seen that, to enhance the di-photon rate, light
doubly-charged particles appear in both scenarios. Concretely, in
scenario A, we have a light doubly-charged triplinos $\chi^{++}$. In
scenario B, besides a light $\chi^{++}$, a light doubly-charged
Higgs boson $H^{++}$ around 100 GeV is predicted. In this
subsection, as a preliminary discussion, we will briefly comment on
their prospects for detecting but leave more quantitative studies
elsewhere.

\subsubsection{Doubly-charged triplino}

At the LHC, $\chi^{++}$ can be pair produced via
the Drell-Yan process or associated produced with a singly-charged
triplino. For $\chi^{++}$ around 100 GeV, the production cross
section can be as large as $10$ pb (It quickly drops to a few 0.1 pb
for a 200 GeV $\chi^{++}$)~\cite{Demir:2008wt}. Moreover, the singly-
and doubly-charged triplinos decay produce multi $W-$bosons, through
\begin{comment}
 which
can be seen from the kinematic term for the fermionic triplets
\begin{align}\label{}
{\cal L}=\f{g_2}{2}{\rm Tr}\L \wt T_u^\dagger\sigma^\mu[W_\mu,\wt
T_u]\R+ \f{g_2}{2}{\rm Tr}\L\wt  T_d^\dagger\sigma^\mu[W_\mu,\wt
T_d]\R.
\end{align}
\end{comment}
the typical decay topologies
\begin{align}\label{}
\chi^+_1 \ra \chi_1^0 W^+,\quad \chi^{++}\ra \chi^+_1 W^+\ra
\chi_1^0W^+W^+.
\end{align}
Some of the $W-$bosons may be off-shell, depending on the mass
splitting among $\chi^{++}$, $\chi_1^+$ and $\chi_1^0$. If
$m_{\chi^{++}}-m_{\chi_0}<m_W$, both $W$ bosons are off-shell. It
is likely to be the case in scenario B where $T_{u,d}$ do not
significantly couple to the Higgs doublets and thus mixing (or loop corrections)
induced splitting is only at the GeV order. As a consequence, the
resulting leptons are too soft to be detected and
$\chi^{++}(\chi^+)$, if not long-lived, behaves as missing energy at
the LHC. Such a pessimistic situation is not of interest.

We only consider cases with sufficiently large mass splittings.
Scenario A is an example, where we typically have such a mass spectrum:
${\chi_1^+}$ has mass around 100 GeV and $m_{\chi^{++}}=\mu_T\sim150-300$ GeV
(see Fig.~\ref{diphoton}), while the LSP $\chi_1^0$ has an even smaller mass,
e.g., 80 GeV. Now consider $\chi^{\pm\pm}$ pair-production, each $\chi^{++}$
decay produces missing energy and energetic same-sign leptons.
Then the tetraleptons plus missing energy signature
\begin{align}\label{}
pp\ra \chi^{++}\chi^{--}\ra
(\ell^+_1\ell^+_2)(\ell_3^-\ell^-_4)E_T^{\rm miss}
\end{align}
has extremely suppressed backgrounds, say $\sim 10^{-3}$ fb~\cite{Demir:2008wt},
and thus has a very promising discovery prospect.

\subsubsection{Doubly-charged Higgs boson}

Scenario B predicts $m_{H^{++}}\simeq100$ GeV, which is of interest.
Recalling that even we work in the supersymmetric type II seesaw,
extremely tiny couplings $(\ld_L)_{ij}\sim{\cal O}(10^{-10})$ are
required. As a result, the di-boson mode $H^{++}\ra W^+W^{+}$ is
dominant in $H^{++}$ decays~\footnote{Note that the decay channel
$H^{++}\ra H^+W^+$ with a real $H^+$ is forbidden, since in the
triplets $T_d$, the doubly-charged component is the lightest one.
The reason can be explicitly found in Eq.~(\ref{hdoubly}), where
$T_d^{++}$ receives a negative mass term from the coupling to the
Higgs doublets while $T_d^{+}$, by contrast, gets a decrease. In
other words, such a mass order is correlated with the fact that
$T_d^{++}$ enhancing the di-photon rate while $T_d^+$ acts
oppositely.}. Such a light charged particle can be probed at LEP and
the LHC. Before heading towards the main intent of this section, we
show that $H^{++}$ promptly decays at collidrs. Concretely, its
width can be estimated in the massless limit:
\begin{align}\label{}
    \Gamma(H^{++}\ra W^+ff') \sim& \f{g_2^6 }{288\pi^3 }\f{v_{T_d}^2
    m_{H^{++}}^3}{m_W^4}.
\end{align}
Taking $m_{H^{++}}=100$ GeV and $v_{T_d}=1$ GeV, the resulted flying
distance of the decaying $H^{++}$ is $\sim$10$^{-7}$ mm.
Thus, even taking into consideration the possible large phase space
suppression, it is found that $H^{++}$ still promptly decays at colliders.

We now examine whether it has been excluded or not by LEP and/or
LHC. The LEP II searches for the tetraleptons signature
$\ell^+\ell^+\ell^-\ell^-$~\cite{LEP} can be used to constrain
$H^{++}$ with di-boson decay:
\begin{align}\label{}
  e^+e^-\ra H^{--}  H^{++}\ra W^-W^{-*}W^+ W^{+*}.
\end{align}
But the cross section of tetraleptons signature is suppressed by the
fourth power of branching ratio of $W$ leptonic decay, and the LEP
II bound is thus weak~\cite{LEP}. Similarly, at the LHC, the CMS and
ATLAS same-sign lepton~\cite{Han:2007bk,Melfo:2011nx} as well as
multi-lepton plus missing energy searches only place weak bounds on
$m_{H^{++}}$. Actually, in the light of a recent
study~\cite{Kanemura:2013vxa}, the present strongest bound is placed
by the 7 TeV LHC with integrated luminosity 4.7 fb$^{-1}$ searching
same-sign lepton~\cite{ATLAS:SL}, which gives a quite loose lower
bound: $m_{H^{++}}>60$ GeV. As the integrated luminosity accumulates
to 20 fb$^{-1}$, the lower bound reaches 85 GeV. In summary, the
current collider search results still allow a large parameter space
for a light doubly-charged Higgs boson. Then, how to probe, discover
and reconstruct such a particle at the future LHC is of interest,
says via considering two hadronic $W-$bosons which produce
signatures like
\begin{align}\label{}
pp\ra W^-W^{-*}W^+ W^{+*}\ra(jjjj)(\ell_1^-\ell^-_2)E_T^{\rm miss}.
\end{align}
We leave it in a specific work.

\begin{comment}Ref.~\cite{Han:2007bk} also have considered same-sign di$-W$
search from doubly-charged Higgs, using the signature
\begin{align}\label{}
pp\ra \chi^{++}\chi^{--}\ra
(\ell^+_1\ell^+_2)(\ell_3^-\ell^-_4)E_T^{\rm miss},\quad {\rm
or}\quad (jjjj)(\ell_3^-\ell^-_4)E_T^{\rm miss}
\end{align}
But it only gave a discussion on the mass region above 200 GeV. In
our interesting mass region, the resulted leptons or jets tend to be
soft. But its production rate is quite considerable and the pair
produced $H^{++}$ can be significantly boosted, so we may use the
fat-jets technique to dig out this doubly-charged Higgs boson. We
leave this promising scenario for future work.
\end{comment}

\section{Discussion and conclusion}

Inspired by the relatively heavy Higgs boson mass and hints for
the Higgs to di-photon excess, we, with guidance for naturalness,
investigate low energy SUSY including light triplets with hypercharge
$\pm1$. It is found that the TSSM shows two attractive scenarios:
\begin{itemize}
  \item One is the large $\ld_{u}$ and $\ld_d$ scenario, which
  is able to not only enhance $m_h$ but also di-photon rate.
  We point out that there is a tension between these two
  aspects, i.e., the latter tends to render the former excessively
  enhanced. Fortunately, the doublet-triplet mixing effects can substantially
  reduce $m_h$ to the desired value.
  \item Oppositely, the other one is characterized by
  negligibly small $\ld_{u,d}$, which, despite of failing enhancing
  $m_h$, can provide an elegant explanation to the origin of the
  di-photon excess. On top
  of that, it provides an inelastic non-thermal DM
  candidate, i.e., the neutral pseudo Dirac triplino.
\end{itemize}
Both scenarios predict light doubly-charged objects, and thus we
made a preliminary analysis of their discovery prospect at the LHC,
based on the signatures from same-sign $W-$bosons.

We would like to end up this paper by commenting on the grand
unification prospect. In the $SU(5)$-GUT, triplets $T_{u,d}$
come from the symmetric rank two tensors of $SU(5)$. Under the
decomposition $SU(5)\ra S(3)_C\times
SU(2)_L\times U(1)_Y$, we have $15=(1,3,1)\oplus(3,2,\f{1}{6})\oplus
(6,1,-\f{2}{3})$ and similarly for $\overline{15}$. Adding these light
particles renders the gauge couplings
non-perturtative below the GUT-scale. But interestingly it is consistent with
scenario A, where the Landau pole may be hit below the GUT-scale
and thus non-perturtative unification is required~\cite{Delgado:2012sm}.
As for scenario B, introducing magic fields can lead to the gauge coupling
perturtative unification~\cite{FileviezPerez:2012ab}.

\begin{comment}
To end this section, we make a comment  the implication  So in
general light colored states are expected, then constraint from
proton decay and collider requires consideration. Fortunately, they
do not give rise to rapid proton decay, since the lepton-quark from
$\overline{15}$ that contains $\bar T$ does not couples to matters
(it is different to the Higgs triplets relating with Higgs doublet,
where  both states couple to matters).
 The collider constraint only emerges   from direct production
 associated with  gauge boson because of the smallness of Yukawa coupling
  $\ld_{L}$. The colorful states are charged, thus LEP imposing model independent bound.

\section*{Acknowledgement}

We would like to thank Ran Huo, Chunli Tong and Lilin Yang, very
much for helpful discussions,
 and thank Tao Liu very much for the collaboration in the early stage of this project.
This research was supported in part by the Natural Science
Foundation of China under grant numbers 10821504, 11075194,
11135003, and 11275246, and by the DOE grant DE-FG03-95-Er-40917
(TL).
\end{comment}

\section*{Note added}

After the completion of this work, we noticed
that~\cite{Chun:2013ft} appeared on the arxiv. This paper, working
in the supersymmetric type II seesaw model and focusing on the
doubly-charged Higgs boson as the source of di-photon excess as well
as its LHC implication, overlaps with a part of our discussion. Our
results agree with each other.

\appendix

\section{Yukawa couplings' RGEs}\label{RGEs}

For using in the text, here we present the RGE running of the new
Yukawa couplings $\ld_{u,d}$ as well as the running of relevant
parameters, e.g., the Yukawa couplings $h_t$,$h_b$ and $h_\tau$.
\begin{eqnarray}
16\pi^2\frac{d\lambda_u}{dt} &=& \lambda_u\left[ 6
h_{t}^{\dagger}h_t  + 14 \lambda_u^\dagger \lambda_u -7 g_2^2 -
\frac{9}{5}g_1^2 \right],\cr
 16\pi^2\frac{d\lambda_d}{dt}&=& \lambda_d\left[ 6 h_{b}^{\dagger}h_b  +
 2 h_{\tau}^\dagger h_\tau  +              14 \lambda_d^\dagger
\lambda_d -7 g_2^2 - \frac{9}{5}g_1^2 \right],\cr 16\pi^2
\frac{dh_t}{dt} &=& h_t \left[ 6 h_t^\dagger h_t  + h_b^\dagger h_b
+6\lambda_u^\dagger \lambda_u -\frac{16}{3}g_3^2 -3g_2^2 -
\frac{13}{15}g_1^2 \right] ,\cr 16\pi^2 \frac{dh_b}{dt} &=& h_b
\left[  6 h_b^\dagger h_b + h_\tau^\dagger h_\tau +h_t^\dagger h_t
+6 \lambda_d^\dagger \lambda_d - \frac{16}{3}g_3^2 -3g_2^2 -
\frac{7}{15}g_1^2     \right],\cr 16 \pi^2 \frac{dh_\tau}{dt} &=&
h_\tau\left[ 3 h_b^\dagger h_b + 4h_\tau^\dagger h_\tau+6
\lambda_d^\dagger \lambda_d -3g_2^2 -\frac{9}{5}g_1^2\right].
\end{eqnarray}
here $t = \ln\frac{\mu}{\mu_0}$ and $\mu$ is the running scale. The
presence of light triplets at the weak scale affects the MSSM gauge
coupling runnings:
\begin{eqnarray}
16\pi^2 \frac{dg_1}{dt} & =& \frac{51}{5}g_1^3,\cr 16\pi^2
\frac{dg_2}{dt} & =&  5g_2^3,\cr
 16\pi^2 \frac{dg_3}{dt} &
=& -3g_3^3.
\end{eqnarray}
As expected, they will not show unification.

\section{The Higgs potential and mass matrix}

The total Higgs potential contains three parts, among which the
$F-$term and soft term can be obtained in the text, while the
$D-$terms with respect to the $SU(2)_L\times U(1)_Y$ groups are
given by
\begin{align}\label{Dterms}
V_D=&\f{g_1^2}{2}\left({\rm Tr}(-T_u^\dagger T_u+T_d^\dagger T_d)
+\f{1}{2}\L H_u^\dagger  H_u-  H_d^\dagger  H_d \R \right)^2+ \cr
&\f{g_2^2}{2}\sum_{a=1,2,3}\left[ \f{1}{2}{\rm
Tr}(T_u^\dagger[\sigma^a,T_u])+\f{1}{2}{\rm
Tr}(T_d^\dagger[\sigma^a,T_d]) +\f{1}{2}\L H_u^\dagger\sigma^a H_u-
H_d^\dagger\sigma^a H_d \R \right]^2.
\end{align}
Collecting all the terms and adopting in the fields decomposing as
done in Eq.~(\ref{GSB:basis}), we get the CP-even Higgs mass square
matrix ${M}_{S}^2$ with entries (in the basis $(h_1,h_2,h_3,h_4)$)
\begin{align}\label{ms2}
({M}_{S}^2)_{11}\doteq&M_A^2+m_Z^2\left(1+\f{\ld_u^2+\ld_d^2}{g^2}\right)
\sin^22\beta,\cr
({M}_{S}^2)_{12}\doteq&-m_Z^2\left[\cos2\beta\L1+\f{\ld_u^2+\ld_d^2}{g^2}\R-\f{\ld_u^2-\ld_d^2}{g^2}\right]\sin2\beta,\cr
({M}_{S}^2)_{13}\doteq&m_Z \left[2 \ld_u \mu \cos2\beta/g - (A_u +
\ld_d\mu_T )\sin2\beta/g\right],\cr ({M}_{S}^2)_{14}\doteq&m_Z
\left[2\ld_d \mu \cos2\beta/g +(A_d + \ld_u\mu_T
)\sin2\beta/g\right],\cr
({M}_{S}^2)_{22}=&m_Z^2\left[\cos^22\beta+\f{4}{g^2}\L\ld_d^2\cos^4\beta+\ld_u^2\sin^4\beta\R\right]\cr
({M}_{S}^2)_{23}\doteq&-2m_Z {\cal M}_{u}/g,\quad
({M}_{S}^2)_{24}\doteq-2m_Z {\cal M}_{d}/g,\cr
({M}_{S}^2)_{33}\doteq&m_Z^2 {\cal M}_{u},\quad
({M}_{S}^2)_{34}\doteq B\mu_T,\quad ({M}_{S}^2)_{44}\doteq m_Z^2
{\cal M}_{d},
\end{align}
where $M_A^2\equiv 2B\mu/\sin2\beta$. In reality, in the heavy
$m_{T_u}^2$ limit (similarly applied to $T_d^0$), $({M}_{S}^2)_{33}$
can be simply written in a more clear way after using
Eq.~(\ref{vTud}),
\begin{align}\label{ms2}
({M}_{S}^2)_{33}\doteq&m_{T_u}^2.
\end{align}

\end{document}